\documentclass{article}
\usepackage{frascatiphys,here,graphicx,subfigure}
\begin{document}
\title{
SEARCHES FOR MULTIBARYON STATES WITH $\Lambda$ HYPERON SYSTEMS IN pA
COLLISION AT 10 GeV/c  }
\author{
P.Zh.Aslanyan      \\
{\em Joint Institute for Nuclear Research, Dubna, Russia
 and Yerevan State of University} \\
V.N.Emelyanenko    \\
{\em Joint Institute for Nuclear Research} }
\maketitle
\baselineskip=11.6pt
\begin{abstract}
Experimental data as a stereo photographs from the 2m propane bubble
chamber LHE, JINR have been analyzed for exotic multibaryon
metastable and stable states searches. A number of peculiarities
were found in the effective mass spectra of: 1)$\Lambda
\pi^{\pm}$,$\Lambda \pi^+ \pi^-$, $\Lambda p$, $\Lambda p p$,
$\Lambda p \pi$,$\Lambda \Lambda $ and $\Lambda K^0_S$ subsystems.
The observed well known $\Sigma^{*+}$(1385),$\Lambda ^*(1600)$ and
$K^{*\pm}$(892)resonances are good tests for this method. The width
of $\Sigma^{*-}(1385)$ for p+A reaction is two time larger than that
presented in PDG. The $\Lambda \pi^-$ spectrum observed enhancement
in mass range of 1345 MeV/$c^2$ which interpreted as a stopped in
nucleus $\Xi^-$. The cross section of stopped $\Xi^-$ production is
$\approx$ 8 times larger than obtained by fritiof model with same
experimental conditions.
\end{abstract}
\baselineskip=14pt
\section{Prewiew}
There are a few  actual problems of  nuclear and particle physics
which are concerning subject of this report\cite{v0}-\cite{hs07}.
These are following: in-medium modification of hadrons, the origin
of hadron masses, the restoration of chiral symmetry, the
confinement of quarks in hadrons, the structure of neutron stars.
Strange multi-baryonic clusters are an  exiting  possibility to
explore the properties of cold dense baryonic matter and
non-perturbative QCD. Multi-quark states, glueballs and hybrids have
been searched for experimentally for a very long time, but none is
established

\section{Experiment}
 The full experimental information of more than 700000 stereo photographs
  are used to select of  events by $V^0$ channel\cite{v0}.The momentum resolution
 charged particles are found to be $<\Delta P/P>=$2.1 \% for stopped  particles and $<\Delta P/P>$=9.8 \%,
 for nonstopped  particles. The mean values of measurement errors for the depth and azimuthal angles are equal
 to $\leq$0.5 degrees. The masses of the identified 8657-events with $\Lambda$ hyperon  4122-events with
$K_s^0$ meson  are consistent with their PDG values\cite{v0}.The
experimental total cross sections are equal to 13.3 and 4.6 mb for
$\Lambda$ and $K_s^0$ production in the p+C collisions at 10 GeV/c.
Protons can be identified by relative ionazation over the following
momentum range: 0.150$< P <$ 0.900 GeV/c.

 The  background has been  obtained by methods: polynomial function,
 mixing angle and by FRITIOF model \cite{lk}. The statistical significance of resonance
peaks were calculated as NP /$\sqrt{NB}$, where NB is the number of
counts in the background under the peak and NP is the number of
counts in the peak above background.

\section{($\Lambda, \pi^+$) and ($\Lambda, \pi^- $) spectra}

\begin{figure}[H]
    \begin{center}
\subfigure[]
        {\includegraphics[width=55mm,height=80mm]{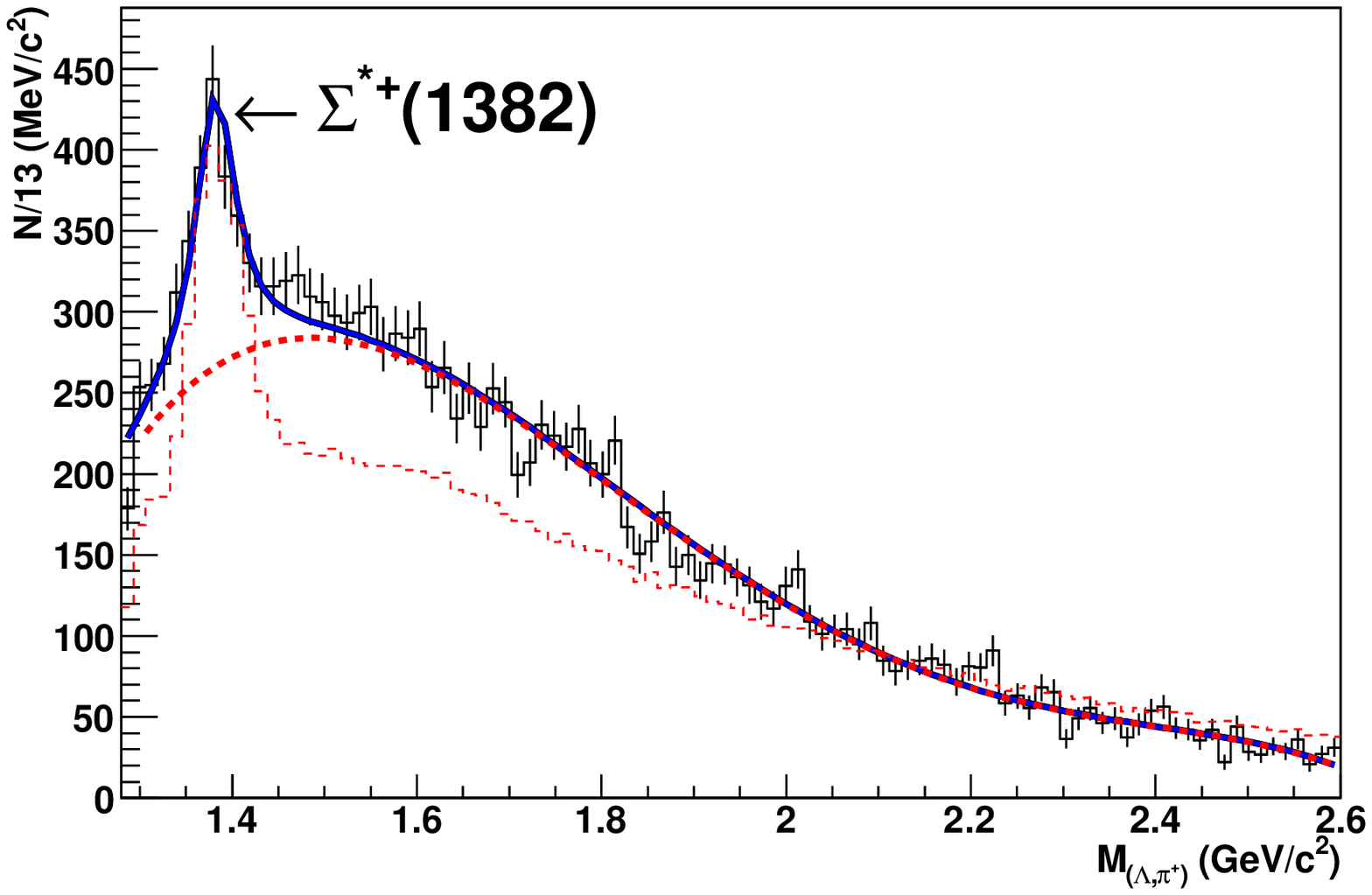}}
\subfigure[]
        {\includegraphics[width=55mm,height=80mm]{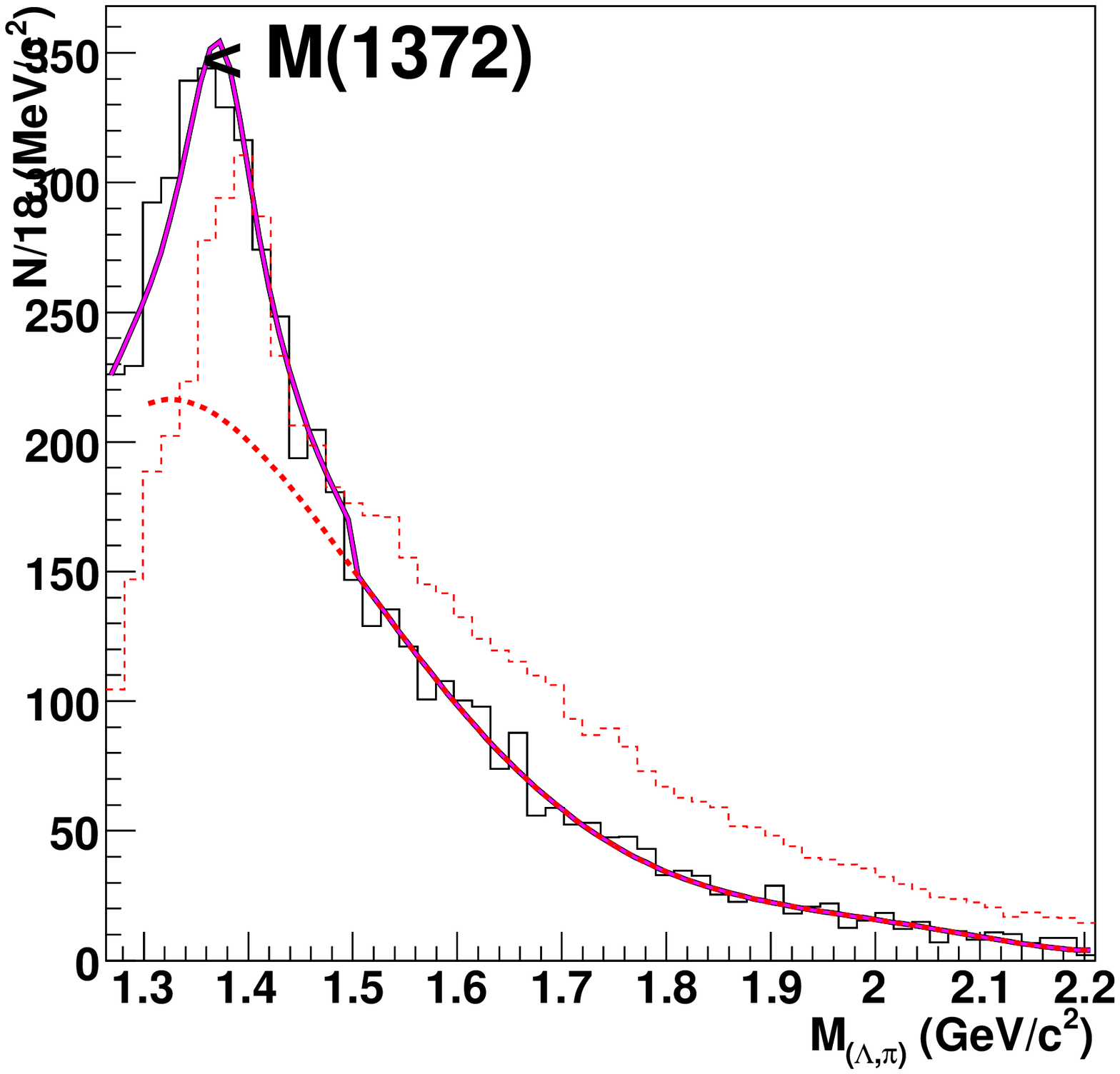}}

        \caption{\it a)The $\Lambda \pi^+$ - spectrum; b)All $\Lambda\pi^-$  comb with
  bin size of 18 MeV/$c^2$.
The simulated events by FRITIOF is the dashed histogram. The
background is the dashed curve.}
\label{lpi}
  \end{center}
\end{figure}

The $\Lambda\pi^+$- effective  mass distribution for  all 15444
combinations with bin size of 13  MeV/$c^2$ in  Fig.\ref{lpi}a has
shown\cite{spin06}-\cite{hs07}. The resonance with similar decay
properties for $\Sigma^{*+}(1382)\to\Lambda \pi^+$ identified which
was a good test for this method. The decay width is equal to $\Gamma
\approx$ 45 MeV/$c^2$. $\Delta M/M =0.7$ in range  of
$\Sigma^{*+}(1382)$ invariant mass.  The cross section of
$\Sigma^{*+}(1382)$ production (540 exp. events) is approximately
equal to 0.9 mb for p+C interaction.

The $\Lambda\pi^-$- effective  mass distribution for  all 6730
combinations with bin sizes of 18 and 12 MeV/$c^2$ in
Fig.\ref{lpi}b,\ref{lp}a has shown. The solid curve(Fig.\ref{lpi}b)
is the sum of the background (by the polynomial method ) and 1
Breit-Wigner resonance($\chi^2/N.D.F.=39/54$). There is significant
enhancement in the mass range of 1372 MeV/$c^2$  with 11.3
S.D.,$\Gamma$ =93 MeV/$c^2$. The cross section of $\Sigma^{*-}$
production ($\approx$680 events) is equal to $\approx$ 1.3 mb at 10
GeV/c for p+C interaction. The width for $\Sigma^{*-}$ observed
$\approx$2 times larger  than PDG value. One of possible explanation
is nuclear medium effects on invariant mass spectra of hadrons
decaying in nuclei\cite{sig}.

Figure \ref{lp}a shows $\Lambda\pi^-$ effective mass distribution
with  bin size of 12 MeV/$c^2$, where there are  significant
enhancements in mass regions of 1345(3.0 S.D.) and 1480(3.2) too.
The solid curve(Fig.\ref{lp}a) is the sum of the background  and 1
Breit-Wigner resonance ($\chi^2/N.D.F.=109/88$). The background
(dashed )curve is the sum of the six -order polynomial  and 1
Breit-Wigner function with parameters for identified resonance
$\Sigma^{*-}$(1385)(Fig.\ref{lpi}b). There are negligible
enhancements in mass regions of 1410, 1520 and 1600 MeV/$c^2$. The
cross section of $\Xi^-$- production ($\approx$60 events) stopped in
nuclear medium is equal to  315 $\mu$b  at 10 GeV/c for p+propane
interaction. The observed number events with $\Xi^-$ by weak decay
channel is equal to 8 (w=1/$e_{\Lambda}$ =5.3, where  is a full
geometrical weight of registered for $\Lambda$s)\cite{H}.Then
experimental cross section   for identified $\Xi^-$ by weak decay
channel\cite{H} is equal to 44$\mu$b and 11.7$\mu$b in p+propane and
p+C collisions, respectively, which are conformed with FRITIOF
calculation. The observed experimental cross section for stopped
$\Xi^-$(60 events) is 8 times larger than the cross section which is
obtained by fritiof model with same experimental conditions. The
width of $\Sigma^{*-}(1385)$ for p+A reaction is two time larger
than that presented in PDG.Figures shows that there is observed
$\Sigma^{*-}$(1480) correlation  which is agreed with report from
SVD2 collaboration too.

 \section{($\Lambda, p$) and ($\Lambda, p, p $) spectra}

\begin{figure}[H]
\begin{center}
\subfigure[]
        {\includegraphics[width=55mm,height=80mm]{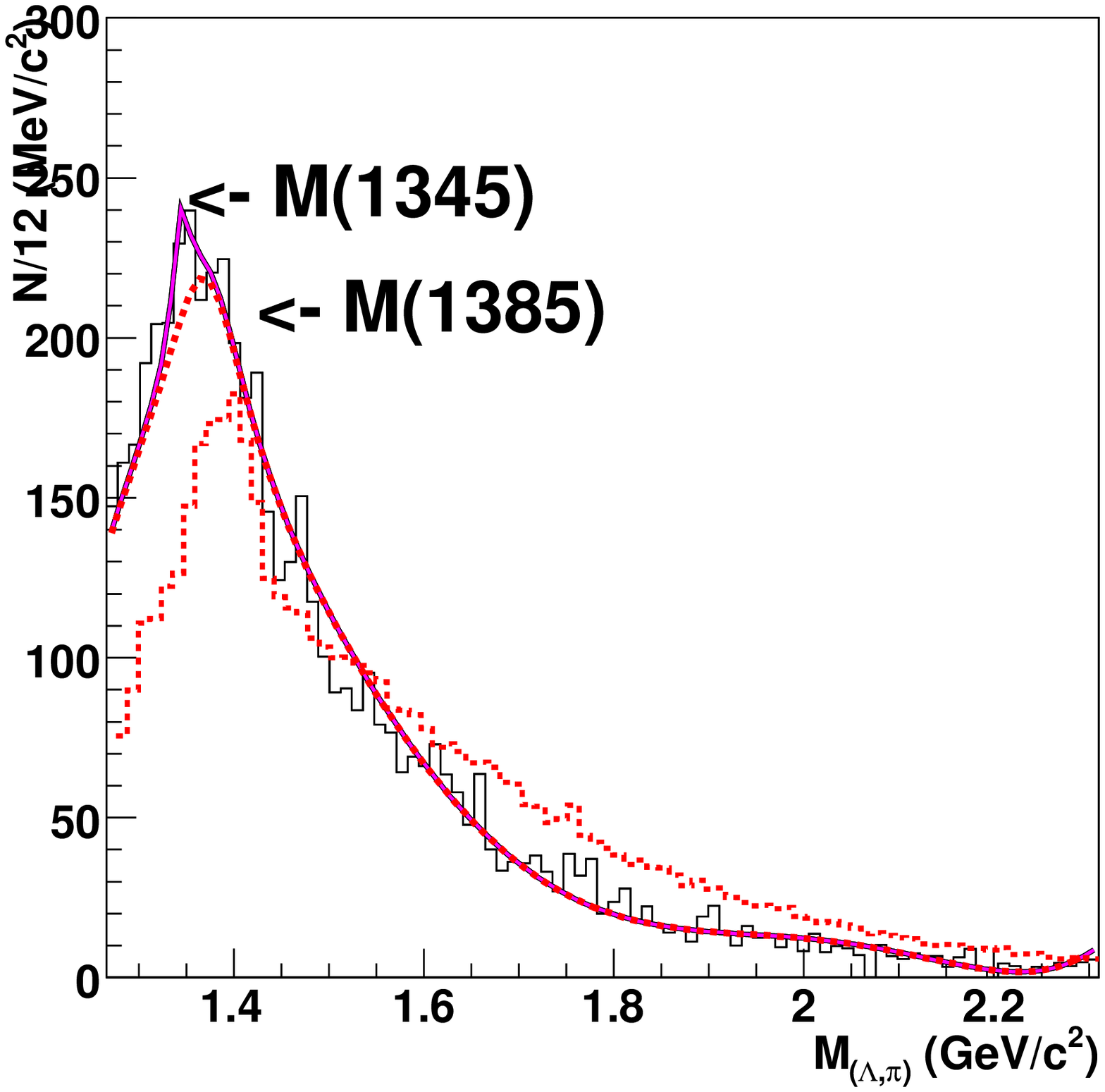}}
\subfigure[]
         {\includegraphics[width=55mm,height=80mm]{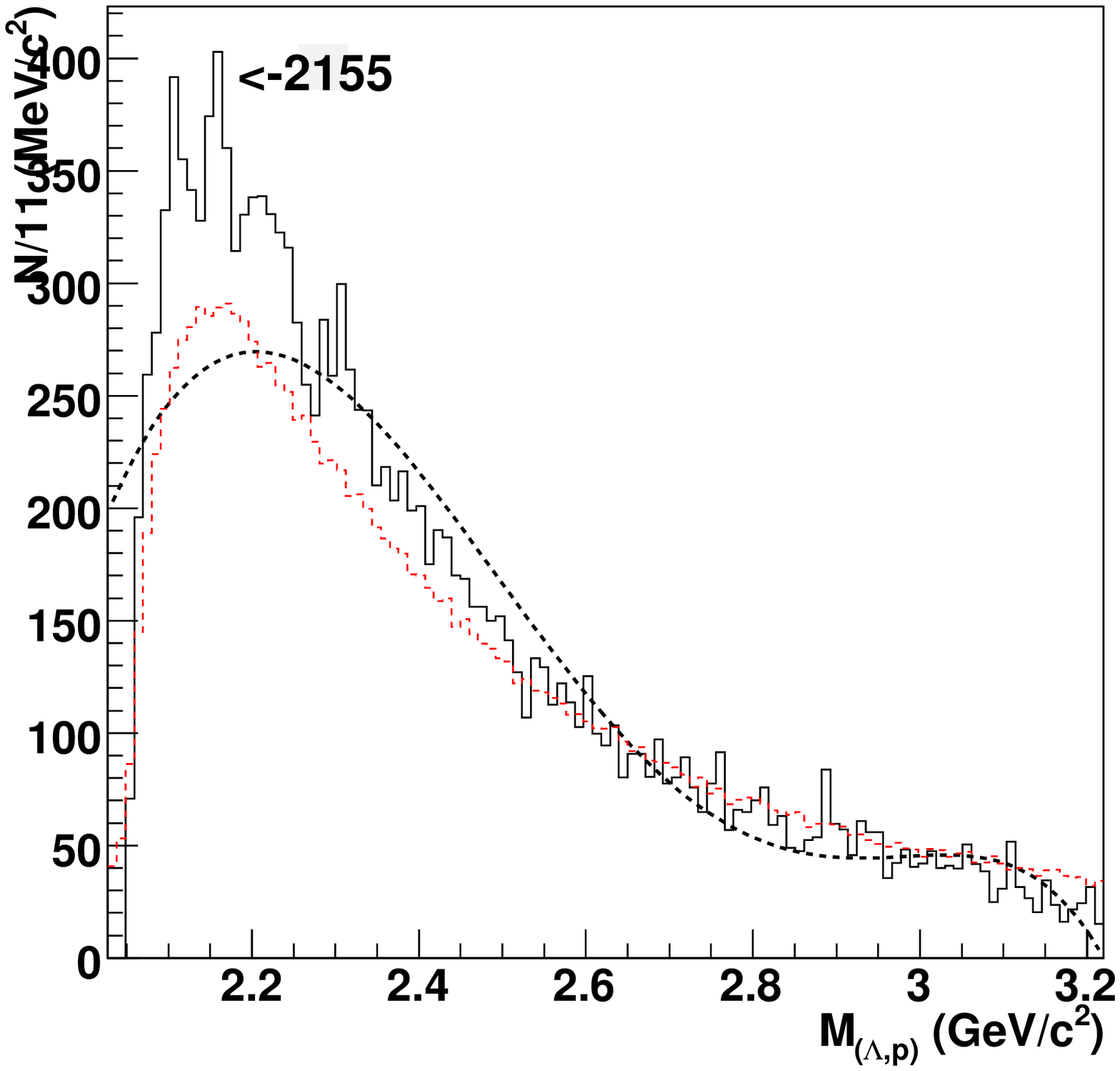}}
\subfigure[]
          {\includegraphics[width=55mm,height=80mm]{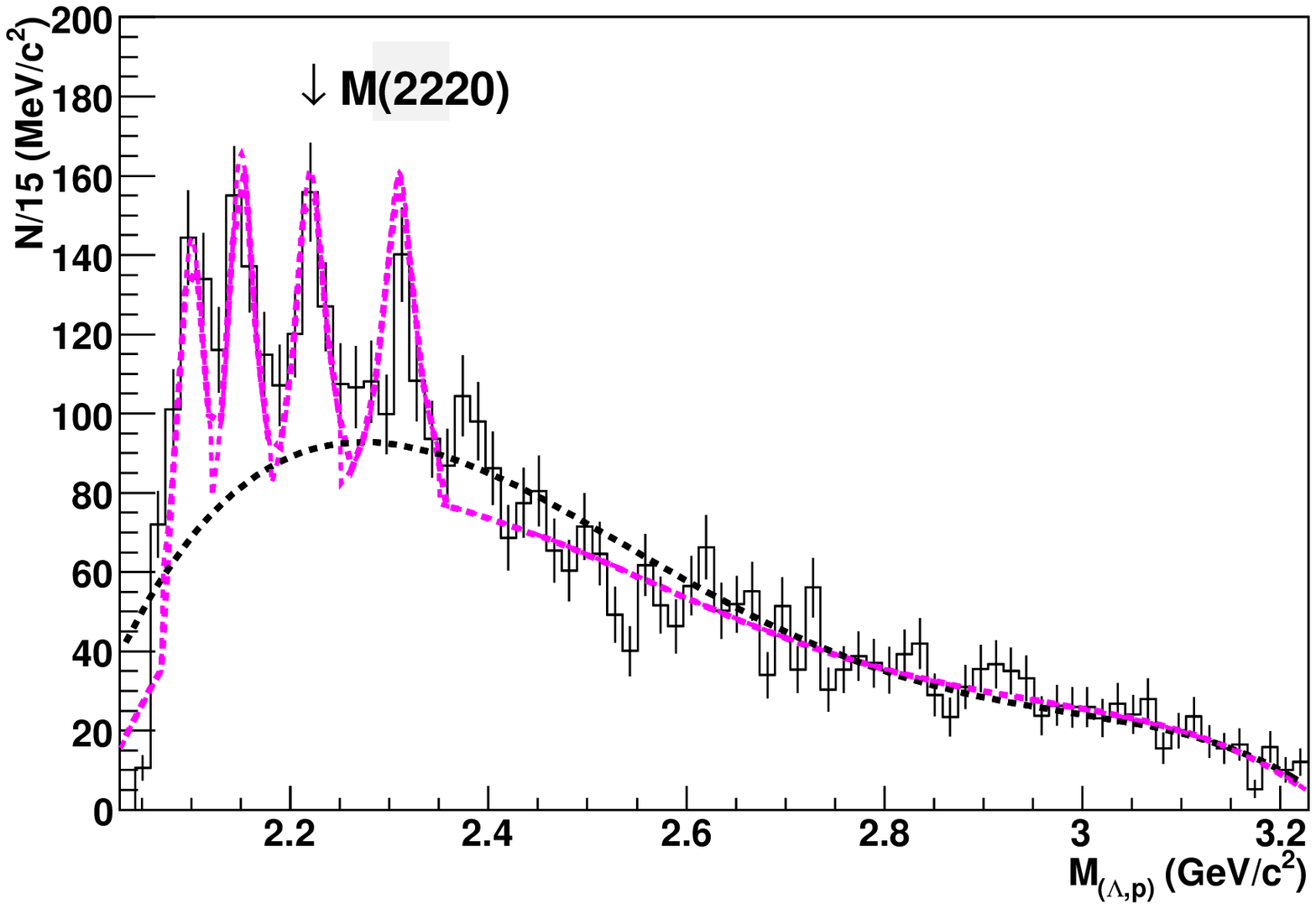}}
\subfigure[]
          {\includegraphics[width=55mm,height=80mm]{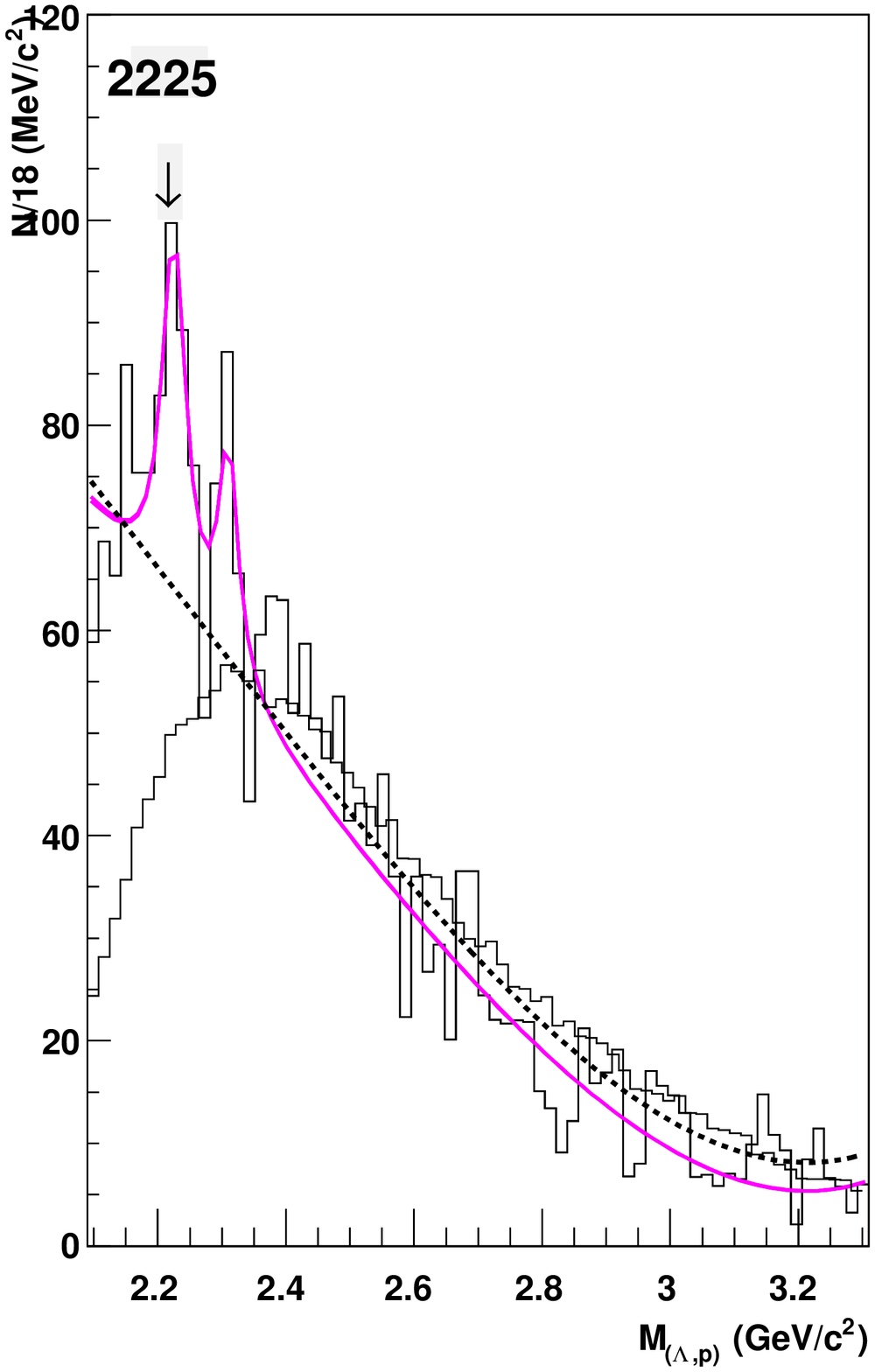}}
   \caption{\it a) $\Lambda \pi^-$spectrum with bin size of 12 MeV/$c^2$.b)All comb for the $\Lambda p$ spectrum;
    c)$\Lambda p$ spectrum with stopped protons in momentum range of 0.14$<P_p<$0.30 GeV/c;d) $\Lambda p$  spectrum for relativistic positive tracks
  in range of $P_p>$1.65 GeV/c. The dashed histogram is simulated events by FRITIOF.}
 \label{lp}
 \end{center}
\end{figure}

 Figure  \ref{lp}b)  shows  the invariant mass  for all $\Lambda
p$ 13103 combinations with bin size of 15 MeV/$c^2 $ (\cite{lp}) .
There are  enhancements in mass regions of 2100, 2150, 2225 and 2353
MeV/$c^2$(Fig.\ref{lp}b).
    There are many published articles\cite{lp}-\cite{hs07}for
    the ($\Lambda p$)invariant mass with identified protons in momentum
    range of 0.350$< P_p<$ 0.900 GeV/c. There are significant enhancements
     in mass regions of 2100, 2175, 2285
 and 2353 MeV/$c^2$.Their excess above background by the
 second method is 6.9, 4.9, 3.8 and 2.9 S.D., respectively. There is also a
 small peak in 2225( 2.2 S.D.) MeV/$c^2$ mass region.

Figure \ref{lp}c  shows the invariant mass of 4011($\Lambda
p$)combinations with bin size 15 MeV/$c^2 $ for stopped protons in
momentum range of 0.14$< P_p<$ 0.30 GeV/c.The  dashed curve is the
sum of the 8-order polynomial  and 4 Breit-Wigner curves with
$\chi^2=30/25$ from fits(Table~\ref{reslp}). A significant peak at
invariant mass 2220 MeV/$c^2$ (6.1 S.D.), $B_K$ ~ 120 MeV was
specially stressed by Professor T. Yamazaki on $\mu$CF2007, Dubna,
June-19-2007 that is conform with KNC model\cite{knc} prediction by
channel of $K^- pp \to \Lambda $p .

 The $\Lambda p$ effective mass distribution for
2025 combinations with relativistic protons over a momentum of P
$>$1.65 GeV/c is shown in Figure \ref{lp}d . The solid curve is the
6-order polynomial function($\chi^2$/n.d.f=205/73).  There are
significant enhancements in mass regions of 2155(2.6 S.D.), 2225(4.7
S.D., with $\Gamma$=23 MeV/$c^2$), 2280(4.2 S.D.), 2363(3.6 S.D.)
and 2650 MeV/c$^2$(3.7 S.D.). These observed peaks for combinations
with relativistic protons P $>$1.65 GeV/c agreed with peaks for
combination with identified protons and  with stopped protons(Table
\ref{reslp}).

\begin{figure}[H]
\begin{center}
\subfigure[] {\includegraphics[width=55mm,height=80mm]{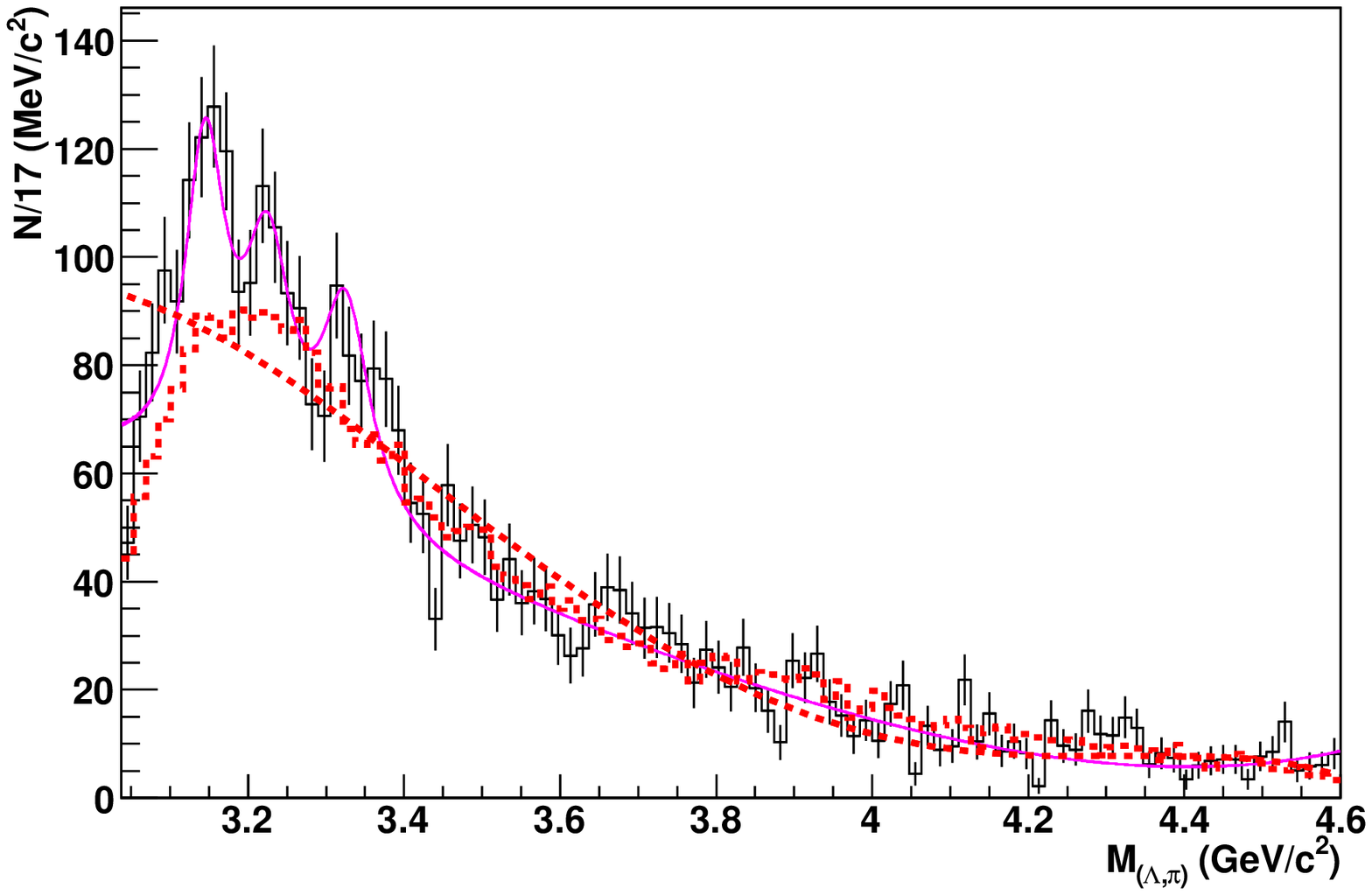}}
\subfigure[] {\includegraphics[width=55mm,height=80mm]{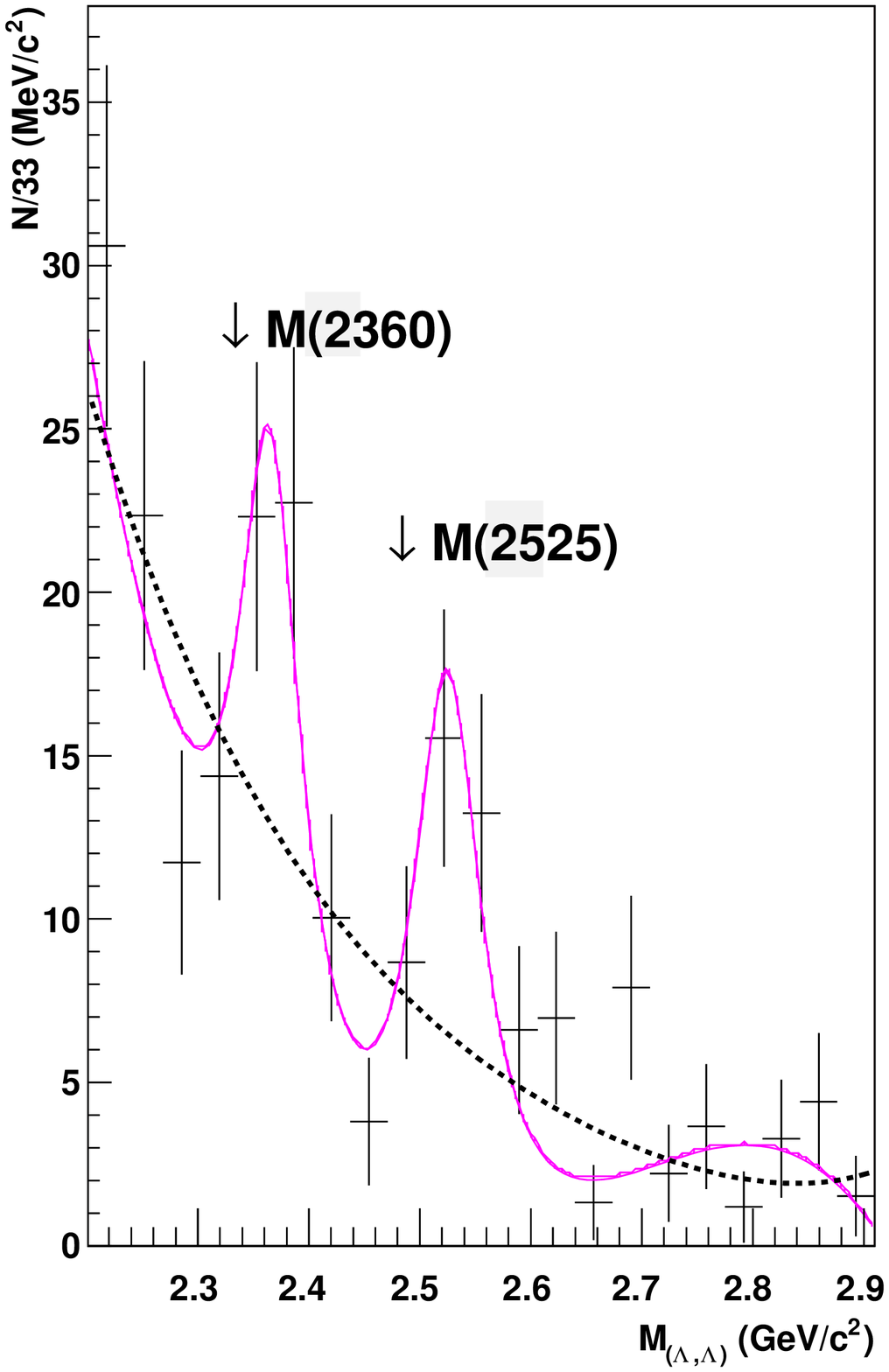}}
\subfigure[] {\includegraphics[width=55mm,height=80mm]{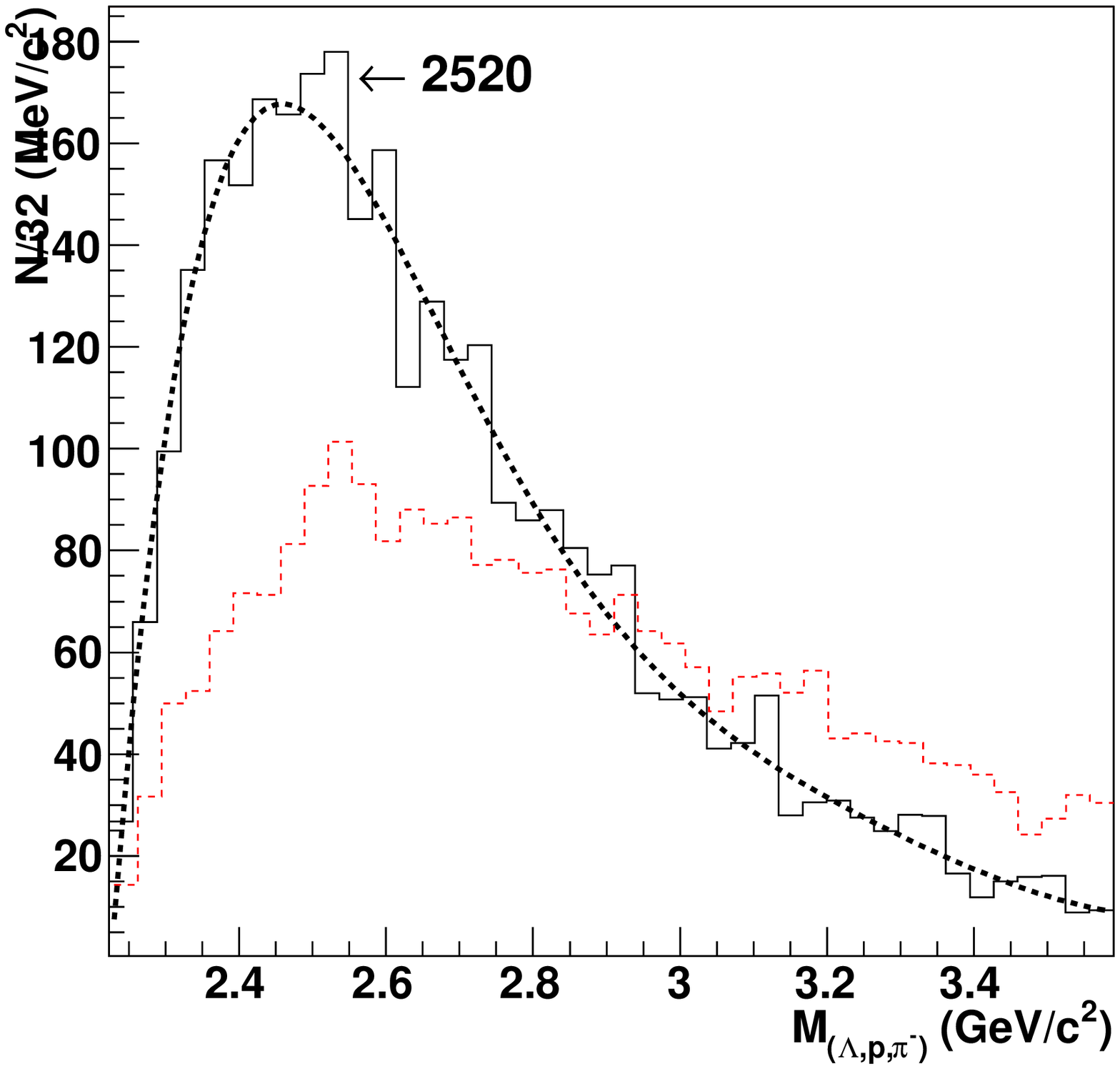}}
\subfigure[] {\includegraphics[width=55mm,height=80mm]{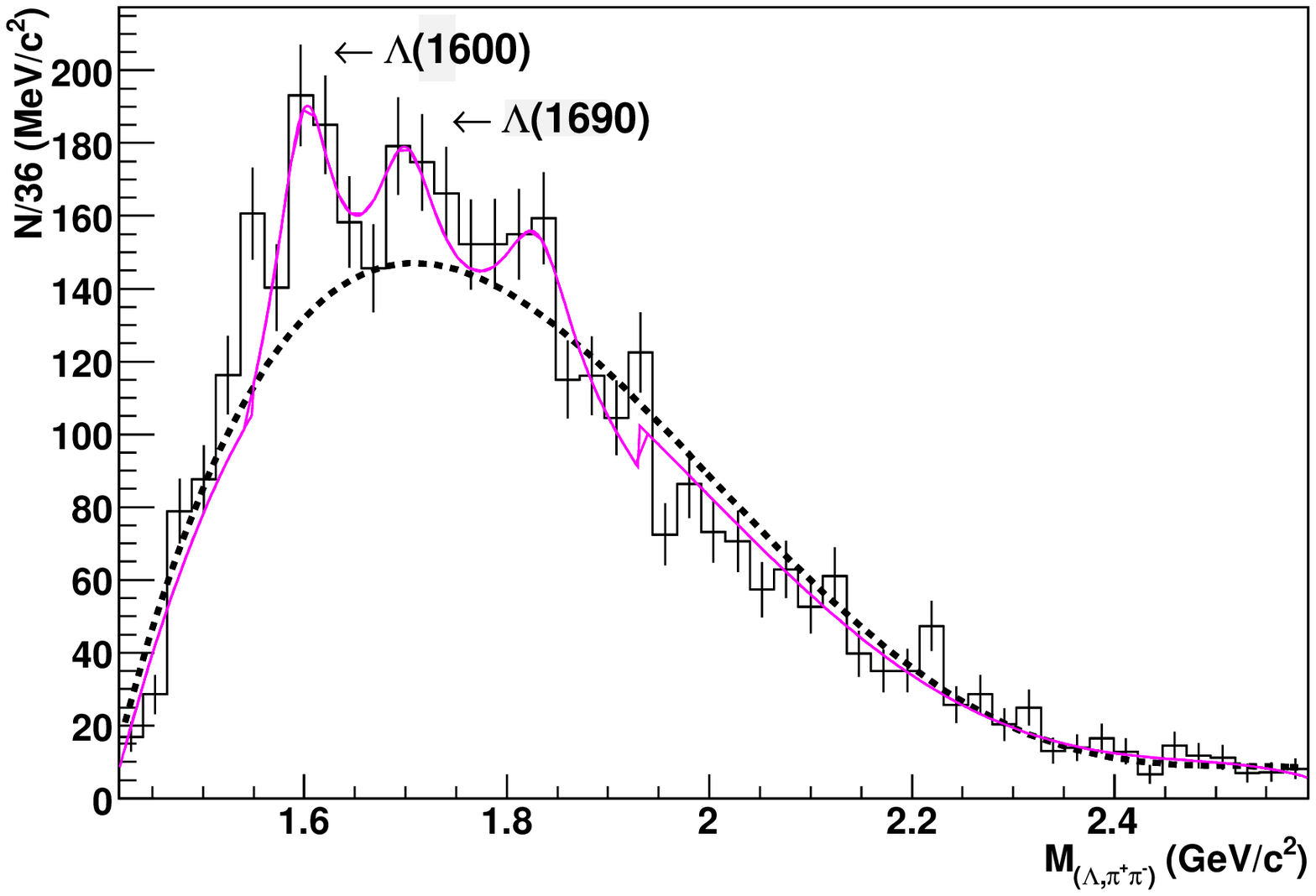}}
  \caption{a)$\Lambda p p$ spectrum with identified
  protons $P_p<$0.9 GeV/c; b)$\Lambda \Lambda$ spectruma)
  c)$\Lambda p \i^-p$ spectrum with identified
  protons $P_p<$0.9 GeV/c; d)$\Lambda \pi^+\pi^-$ spectrum
  for  positive tracks in momentum range of $P_{\pi+}<$0.9 GeV/c. The dashed histogram
  is simulated events by FRITIOF. The experimental background is the dased curve.}
    \label{lpp}
    \end{center}
\end{figure}

The $\Lambda pp$   effective  mass distribution for 3401
combinations for identified protons with a momentum of  $P_p <$0.9
GeV/c is shown in Figure \ref{lpp}a)\cite{spin06}-\cite{hs07}. The
solid curve is the 6-order polynomial
function($\chi^2$/n.d.f=245/58, Fig.\ref{lpp}a ). The backgrounds
for analysis of the experimental data are based on FRITIOF and the
polynomial method. There is significant enhancements in mass regions
of 3145 MeV/$c^2$(6.1 S.D.) and with width 40 MeV/$c^2$. There are
small enhancements in mass regions of 3225(3.3 S.D.), 3325(5.1
S.D.), 3440(3.9 S.D) and 3652MeV/$c^2$(2.6 S.D.)(Table \ref{reslp}).
These peaks from $\Lambda p$ and $\Lambda p p$ spectra were partly
conformed  with experimental results from FOPI(GSI), FINUDA(INFN),
 OBELIX(CERN) and E471(KEK).\\

\section{($\Lambda,\Lambda)$ spectrum}

There is observed significant enhancement in mass region of 2360(4.5
S.D.) Mev/$c^2$ for $\Lambda,\Lambda)$ spectrum in
Figure~\ref{lpp}b)(137 combination). This peak is conformed with
theoretical predictions and with  earlier published result from
neutron exposure by PBC method with very poor statistics too. There
is small enhancement in mass range of 2525 Mev/$c^2$(3.0 S.D.)
too(Table \ref{reslp}).

\section{$(\Lambda,p,\pi^-)$ spectrum}

The ($\Lambda,p,\pi^-$) effective  mass distribution (Fig.
\ref{lpipi}c) for 2975 combinations for identified protons in
momentum range of  P $<$0.9 GeV/c can taken  by the 6-order
polynomial function which is satisfactory described the experimental
data with $\chi^2/(N.D.F.)$=1. But the background by FRITIOF model
do not describe the experimental distribution. The sum of BW (with
mass 2520 MeV/$c^2$ and experimental width 280 MeV/$c^2$) and
FRITIOF model for $\Lambda p\pi^-$  effective  mass distribution is
satisfactory described the experimental data too. Therefore one of
probably interpretation of this peak that it can be reflection from
phase space distribution too. Earlier published result about
observation of resonance with mass 2495 MeV/$c^2$ and width 200
MeV/$c^2$ for $\Lambda p\pi^-$ spectrum by PBC method for neutron
exposure(7 GeV/c) is not uniquely conformed.

\section{($\Lambda,\pi^+,\pi^- $) spectrum}

The $\Lambda\pi^+\pi^-$  effective  mass distribution for  all 3476
combinations with bin size 36 MeV/$c^2$ has shown in Figure
\ref{lpp}d. The dashed curve is the the background by the polynomial
method.There are significant enhancement in mass region of $\Lambda
^*$(1600)(5.5 S.D., $\Gamma_e$=80 MeV/$c^2$,$\Delta M$=25 MeV/$c^2$)
with width 55 (from PDG).  There are small enhancements in mass
regions of $\Lambda^*$(1520)(3.5 S.D.),$\Lambda^*$(1690)(3.8 S.D.)
and $\Lambda^*$(1800)(2.8 S.D.) MeV/$c^2$ which are interpreted as a
reflection from resonances of $\Lambda^*$(1520), $\Lambda^*$(1690)
and $\Lambda^*$(1800) from PDG. There are not observed exotic states
which were earlier observed and published for $\Lambda\pi^+\pi^+$
spectrum (in mass ranges of 1704,2071,2604 Mev/$c^2$)with small
statistic in neutron exposure by PBC method\cite(shakh).

\section{Conclusion}

$\bullet{}$The invariant mass of  $\Lambda\pi^+$ ,
$\Lambda\pi^+\pi^-$ and $K^0_s\pi^\pm$ spectra has observed well
known resonances from PDG as
$\Sigma^{*+}$(1385),$\Lambda^*(1520),\Lambda^*(1600),\Lambda^*(1690)$
and $K^{*\pm}$( 892) which are a good test for this method.\\
$\bullet{}$ A number of important peculiarities were observed  in pA
$\to \Lambda(K^0_S)$ X reactions in the effective mass spectrum for
exotic states with decay modes
(TABLE~\ref{reslp})\cite{lk}-\cite{hs07} :  1) $( \Lambda, \pi), (
\Lambda, \pi^+ ,\pi^-), ( \Lambda,p), ( \Lambda, p, p), (
\Lambda,\Lambda),\Lambda,p ,\pi^-),(\Lambda,K^0_s)$ and
$K^0_s\pi^{\pm}$.\\
$\bullet{}$ Particulary peaks for $(\Lambda,p)$ and $(\Lambda, p,
p)$ spectra are agreed with experimental data from the  reports of
FOPI, E471(KEK), OBELIX, FINUDA collaborations, but there are some
inconstancy by widths.

\section{Acknowledgements}

The work was partly supported by the  grant of RFBR 07-02-08644 and
 org. committee of International Conference HADRON, Frascati,INFN,
 8-12 October, 2007.

\begin{table}[t]
\centering \caption{\it The effective mass , width($\Gamma$) and
 S.D. for observed exotic strange resonances in p+ propane collisions.}
 \vskip 0.1 in
\begin{tabular}{lrrrr}  \hline
$\Lambda p$ &2100&24&5.7\\
 &2150&19&5.7\\
 &2220&23&6.1\\
  &2310&30&3.7\\
   &2380&32&3.5\\
 \hline
 $\Lambda p p$ &3145&40&6.1\\
 &3225&50&3.3\\
 &3325&53&4.8\\
 \hline
 $\Lambda \Lambda$ &2365&55&4.5\\
 &2525&63&3.0\\
 \hline
   $\Lambda K^0_s$&1750&14$\pm$6&5.6\\
 &1795&26$\pm$15&3.3\\
 \hline
 $K^0_s\pi^{\pm}$&890&50&6.0-8.2\\
  &780-800&10&2.5-4.2\\
&720-730&30-125&4.1-15.2\\
&1060&-&7.2\\
 \hline
 \end{tabular}
\label{reslp}
\end{table}

\end{document}